# Enhancement of ultra-thin film emission using a waveguiding active layer

R. Aad*, S. Blaize, A. Bruyant, C. Couteau and G. Lérondel*

*Laboratoire de Nanotechnologie et d'Instrumentation Optique, Institut Charles Delaunay, Université de Technologie de Troyes, 12 rue Marie Curie, BP 2060, 10010 Troyes Cedex, France*

*Corresponding authors: roy.aad@utt.fr, gilles.lerondel@utt.fr;

**ABSTRACT**

We present a theoretical study on the impact of an active optical layer on the emission properties of an ultrathin luminescent film. While the study can be generalized to any material, we focus here on a simple layered medium composed of a conjugated polymers (CPs) thin film, a zinc oxide layer (ZnO) and a sapphire substrate. The study spreads throughout variable aspects including the effect of the structure parameters on the CP luminescence and radiation pattern and more specifically the influence of the absorption and emission properties of the active layer. Comparing between the passive and active layer cases, the obtained results show that an enhancement of the CP luminescence of more than 20 times can be obtained by using an optically active underlying layer. The results can be explained in terms of photon recycling where the optically active layer acts as a photon reservoir and a secondary light source for the ultra thin film. This general concept is of a special interest for ultra-trace chemosensor.

**I. INTRODUCTION**

Fluorescence quenching (FQ) of dyes has been known since 1931, when Kautsky reported FQ by oxygen [1]. However, quenching of conjugated polymers (CPs) is a rather recent development initiated in 1993 [2]. This topic is gaining much interest and receives great attention driven by high expectations in terms of applications and more specifically sensing [3-6]

Our current study concentrates on chemo-sensors. In fact, CPs are great candidates to form the basis for the realization of ultra-trace sensors that profits of their fluorescence quenching feature as a mean for detecting molecules. Moreover, CPs present high quantum and FQ efficiencies that further emphasize their use for sensing applications [7]. Nevertheless, the realization of CPs based sensors encounters difficulties. Actually, the FQ process of the CP is triggered by the capture of an analyte by receiver sites along the polymer chain [8]. Molecule diffusion renders the FQ process highly dependent on the CP layer thickness. As a matter of fact, for less volatile analytes exhibiting low diffusion rates, the CP layer is effectively quenched with surface-bound analytes and the role of diffusion within the film is negligible. As a result, the FQ percentage and the response time of a thick layer would be greatly reduced compared to that of the thin layers [9]. So, it becomes essential to employ thin films in order to enhance the FQ and homogenize the analyte distribution throughout the whole layer thickness and to enhance the sensitivity threshold. On the other hand, in case of an optically pumped CP, an ultra-thin layer becomes inconvenient. Thin layers are weakly excited, less luminescent and more light-squandering compared to thick layers, i.e. presenting low pumping efficiency and weak fluorescence intensity. These are important issues for sensing.

Thus, it is clear that the two needs of enhancing the fluorescence quenching and the luminescence intensity, which are two critical points for the improvement of the detection threshold, cannot be met by the polymer layer alone. In this context, we propose an optical solution where an optically active layer is used to further improve the detection threshold of fluorescent polymers based sensors. The increase of detection threshold of semi-conducting polymer coated on passive substrates have already been studied and discussed in recent papers [10]. However, the influence of an optically active substrate on the luminescence of a coated polymer has never been studied or discussed before. In this paper, we suggest a simple planar optical structure that consists of the CP coated on a zinc oxide (ZnO) layer and we present a theoretical study of the influence on the CP luminescence. The choice of using ZnO as the active layer was made for both its passive and active optical properties that make it a good candidate for ultra violet (UV) photonics [11].

The paper is organized as the following: first, we introduce the optical structure that will be studied. Second, we present the numerical method employed to realize the simulation and then the obtained simulation results are discussed.

## II. OPTICAL STRUCTURE AND MATERIAL ISSUES

In this part, we present the studied optical structure and discuss material issues such as absorption and emission properties.

As shown on fig. 1, the studied structure consists of an ultra-thin luminescent layer (5nm thick CP layer) deposited on an active layer (ZnO) which is grown on a lower refractive index substrate (sapphire). The whole structure is surrounded by air. Thus, the optical structure consists of four dielectric mediums with a planar geometry. The polymer and the zinc oxide in the structure constitute two finite mediums with a fixed $d_{CP}$=5nm and a variable $d_{ZnO}$ thickness respectively, while the air and sapphire constitute two semi-infinite mediums.

In planar structures like the one we proposed, the luminescence is mainly affected by two processes. The first is light interference that is due to the superposition of coherent light waves which either leads to exaltation or destruction of light. The second process is the coupling between the light source and its own light wave that occurs in optical structures where the light emitting source is sandwiched between two interfaces and affects the emission rates of the source [12]. These two processes are directly influenced by the dimensions of the mediums of the structure $d_{CP}$ and $d_{ZnO}$. With the thickness of the CP ($d_{CP}$) being fixed to 5nm and the air and sapphire medium being both semi-infinite, this leaves the thickness of the ZnO layer ($d_{ZnO}$) as the only variable of the structure. Furthermore, the CP and ZnO mediums in our case are optically pumped and photo-excited, which includes another influencing varying parameter that is the angle of excitation ($\theta_{excitation}$).

Besides opto-geometrical parameters, there are material issues that have to be considered to make the structure presented in figure 1 efficient. In fact, if the CP luminescence is to be enhanced by the optical structure, then the CP and the underlying material need to be optically coupled. For that, the optically active material has to meet certain criteria. First, the material needs to be absorbent at the laser excitation wavelength. And second, the material needs to emit light at a wavelength within the CPs absorption band domain.

Actually, the CP used in this study possesses an emission peak at 515nm and an absorption peak at 420nm [13] but can still be, efficiently pumped within a wavelength range starting from 300nm and extending to 475nm.

In this context, the choice of using ZnO as the active layer was made for its optical properties that meet well with the previously mentioned specifications. ZnO is a well-known and well-studied semi-conductor emitting in the UV with an exciton binding energy (60meV) 2.4 times higher than room temperature thermal energy (25meV) what paves the way for an intense near-band-edge excitonic emission at room and higher temperatures. Furthermore, ZnO possesses an absorption band peak (at 325nm) and a luminescence spectra peak (at 375nm) that lie within the CP absorption domain, which makes it a good candidate for an optically active material.

To summarize this section, in the proposed structure, the active layer (ZnO) used as a support for the thin polymer film plays at the same time the role of a reflective media for the polymer emission, an absorber (more than 95% of the excitation is not absorbed by the thin polymer film), an excitation source (by reemitting the absorbed photons), a resonating cavity and a waveguide. At this stage one may wonder of the benefit of using such a structure as compared to metallic layered structures such as a simple metallic mirror. While an increase factor of 4 due to the reflection of the excitation and the emission could be theoretically expected actually, because of efficient non radiative transfer to the metal, the luminescence of the polymer thin film would instead decrease and could be completely quenched. The later effect for extremely thin films in contact with metal is very likely to occur.

Given the complexity of the study, simulation is a necessity for calculating the evolution of the CP luminescence in order to optimize the influencing parameters ($\theta_{excitation}$ and $d_{ZnO}$). For this purpose, we have written and employed a numerical method for calculating the luminescence of the polymer that takes into account all the previously discussed parameters.

## III. NUMERICAL METHOD

The luminescence of the structure was calculated using a numerical method proposed by Stanley et al. named method of source terms [14].

The method of source terms is based on a classical approach to calculate the emission of a thin source plane (sheet) in an arbitrary planar structure; in which, the source plane is modeled by electrical oscillating dipoles. The method considers a weak coupling regime, i.e. the dipole emission is not coherent in space or time. The source is then represented in the calculations by the EM discontinuity [15] as the method does.

To quantify the discontinuity, the method considers three dipole configurations, 2 horizontal dipoles and 1 vertical each distributed on the x, y, and z axes (as sketched on the inset in Fig. 1) respectively. The expression of the EM discontinuity terms (source terms) of each dipole are obtained as given in Table 1. With $k_i$, $k_z$ and $k_{\parallel}$ being the modulus of the wavevector, its projection on the z-axis and its projection on the xy-plane, respectively. *A* presents the source term. The Subscript arrow denotes the propagation direction of the emission: upward arrow for ascending along the z direction and downward for descending along z. The (v) and (h) superscript mark the direction of the emitting dipole whether it is vertical (v) or horizontal (h). And the s and p superscript indicate TE and TM polarization respectively.

Table1. Source Terms for Horizontal and Vertical Dipoles.

| Dipole | Modes | |
|---|---|---|
| | TE | TM |
| Horizontal | $A_{\downarrow,\uparrow}^{(h),s} = \pm\sqrt{\dfrac{3}{16\pi}}$ | $A_{\downarrow,\uparrow}^{(h),p} = \pm\sqrt{\dfrac{3}{8\pi}}\dfrac{k_{z,i}}{k_i}$ |
| Vertical | $A_{\downarrow,\uparrow}^{(v),s} = 0$ | $A_{\downarrow,\uparrow}^{(v),p} = \sqrt{\dfrac{3}{8\pi}}\dfrac{k_{\parallel}}{k_i}$ |

Given the expression of the EM discontinuity terms, the calculation of the emitted power of any arbitrary planar optical structure comes down to solving the following equation system:

$$\begin{pmatrix} E_\uparrow \\ E_\downarrow \end{pmatrix}^{>} - \begin{pmatrix} E_\uparrow \\ E_\downarrow \end{pmatrix}^{<} = \begin{pmatrix} A_\uparrow \\ A_\downarrow \end{pmatrix} \quad (3.1)$$

$$\begin{bmatrix} a_{11} & a_{12} \\ a_{21} & a_{22} \end{bmatrix} \times \begin{pmatrix} 0 \\ E_{ext} \end{pmatrix} = \begin{pmatrix} E_\uparrow \\ E_\downarrow \end{pmatrix}^{<} \quad (3.2)$$

$$\begin{bmatrix} b_{11} & b_{12} \\ b_{21} & b_{22} \end{bmatrix} \times \begin{pmatrix} E_{sub} \\ 0 \end{pmatrix} = \begin{pmatrix} E_\uparrow \\ E_\downarrow \end{pmatrix}^{>} \quad (3.3)$$

Where the 2x2 transfer matrices [16] (a) and (b) respectively depict the propagation to the source from the left and from the right, as sketched in Fig. 1. $E_{ext}$ and $E_{sub}$ indicate the electrical fields emitted respectively in the superstrate and substrate. $E_\uparrow$ and $E_\downarrow$ respectively present the electrical field propagating along and opposite to the z direction. While, the > and < symbols denote whether the field is emitted at the lower or upper side of the source as depicted in Fig. 1.

The outside power is calculated by the following expression that takes into account for the change in solid angle and the Poynting vector flux:

$$\Pi_{out} = \frac{dP}{d\Omega \times ds} = |E_{out}|^2 \frac{n_{out} \times k_{z,out}^2}{n_1 \times k_{z,1}^2} \qquad (3.4)$$

Where $E_{out}$, $n_{out}$ and $k_{z,out}$ respectively present the emitted electrical field, the refractive index and the z-projection of the wavevector in one of the outside mediums (e.g. air or sapphire). While $n_1$ and $k_{z,1}$ present the refractive index and the z-projection of the wavevector in the dipole medium.

When dealing with sources of a finite volume, the method can be generalized by dividing the volume into thin layers (sheets) of dipoles and summing the emitted power from discrete sub-sources, with the weighting coefficients.

Furthermore, since the optical structure is optically pumped, we need to take into account the influence of the structure parameters ($d_{ZnO}$) and the angle of excitation ($\theta_{excitation}$) on the photo-excitation of both CP and ZnO. To do so, we have represented the photo-excitation factor by a weighting coefficient that we calculate for each of the source planes using expression (3.5) which is the expression for dipolar dissipation [17].

$$Coeff_{excitation} = I_D = \frac{1}{2} \times \omega \times \varepsilon_0 \times \chi" \times |E_{local}|^2 \times \Delta x \qquad (3.5)$$

In this expression $\omega$, $\varepsilon_0$ and $\Delta x$ are respectively the angular frequency, the permittivity in vacuum and the source plane thickness, while $\chi"$ and $E_{local}$ are the imaginary part of the electrical susceptibility and the amplitude of the electrical field at the source plane respectively. Because the method models the source as an electrical dipole and since the energy is conserved, the emitted light energy is equal to the energy absorbed (dissipated) by the dipole times the quantum efficiency ($I_{emitted}$=Quantum efficiency × $I_D$). Here simplicity the quantum efficiency was taken to be equal to 1.

Considering that for each of the 3 dipole orientations the oscillating electrical dipole is uniquely excited by an electric field oscillating parallel to its orientation. We distinguish then, a weighting coefficient for each of the 3 dipoles. Table 2 presents the different $|E_{local}|^2$ expressions for the dipoles, knowing that the rest of the weighting coefficient expression is a constant depending of the material and the simulation. In Table 2, $E_\uparrow$ and $E_\downarrow$ respectively present the electrical fields at the source plane ascending and descending along z. while $k_x$, $k_z$, (v), (h), s and p hold the same meaning as previously.

Table 2. Weighting coefficients for Horizontal and Vertical Dipoles.

| Dipole | Modes | |
|---|---|---|
| | TE | TM |
| Horizontal | $\left|E_\uparrow^{(h),s}\exp(ik_z z)+E_\downarrow^{(h),s}\exp(-ik_z z)\right|^2$ | $\left|k_z\times\left\{E_\uparrow^{(h),p}\exp(ik_z z)+E_\downarrow^{(h),p}\exp(-ik_z z)\right\}\right|^2$ |
| Vertical | 0 | $\left|k_x\times\left\{E_\downarrow^{(v),p}\exp(-ik_z z)-E_\uparrow^{(v),p}\exp(ik_z z)\right\}\right|^2$ |

**IV. RESULTS AND DISCUSSION**

We now present the obtained simulation results. In our calculations, we take into account the dispersion of the indices of refraction and the losses of each material. For the optical constant of the ZnO, we took the values of article [18] as a reference, while the values of the CP optical constant were determined by ellipsometry. Table 3 presents the complex refractive indices of the polymer, ZnO and sapphire at 325 (Maximum ZnO absorption), 375 (ZnO emission) and 520nm (polymer emission). Because of the Wurtzite crystallographic structure of ZnO the emission is mainly in the layer plane (perpendicular to the c axis). To account for such anisotropy, the dipole emission was weighted 3 to 1 for TE and TM polarizations respectively. [19]

Two cases were here considered. In the first case, a passive ZnO layer is studied. So, the CP is only excited by the laser excitation. In the second case, the ZnO layer is considered as active. In this case, the CP layer is excited by both the ZnO emission and the laser. The laser excitation is considered at a fixed wavelength of 325nm.

Table 3. Complex refractive indices used in simulation of the polymer, ZnO and sapphire.

|  | $\lambda=325nm$ | $\lambda=375nm$ | $\lambda=520nm$ |
|---|---|---|---|
| Polymer | 1.46+0.021i | 1.46+0.029i | 1.49+0.0002i |
| ZnO | 2.03+0.413i | 2.46+0.310i | 2.03 |
| Sapphire | 1.8 | 1.79 | 1.77 |

A. **Parametrical study: Influence of $\theta_{excitation}$**

Figures 2 shows a 3-dimensional plot of the evolution of the CP total emitted power (TEP, which is the sum of the TE and TM modes) in both air and sapphire medium as a function of the parameters $\theta_{excitation}$ and $d_{ZnO}$. We remind that the CP layer here is both excited from the laser and the ZnO layer. The total emitted power is calculated by integrating the emitted radiation patterns over the whole angular range of emission using the following formula:

$$P_{extracted} = 2\pi \int_0^{\frac{\pi}{2}} \Pi_{out}(\theta) \times \sin(\theta) \times d\theta \qquad (4.1)$$

In this part we are only interested by the influence of $\theta_{excitation}$, where we will further discuss in detail the influence of $d_{ZnO}$ in the upcoming parts. For the behavior of the TEP along with $\theta_{excitation}$, we remark a decrease of the TEP when increasing $\theta_{excitation}$. This is due to the reflectivity of the optical structure that increases with the excitation angle Since in our study we are interested in optimizing the emission, the study of the luminescence at great angles of incidence does not offer much importance. For that, we will limit the study in the succeeding parts to a value of $\theta_{excitation}$ that is null. $d_{ZnO}$ becomes the only influencing parameter.

**B.     Parametrical study: Passive ZnO layer**

In this section, we studied the influence of a passive ZnO layer on the CP luminescence. In this case the CP luminescence is solely excited by the laser excitation.

Figure 3 presents the TEP in air (left) and sapphire (right) for TM (a, c) and TE (b, d) modes. First, we remark a common behavior between the 4 graphs, where the intensity first begins by decreasing before subsequently increasing and oscillating as $d_{ZnO}$ increases. The TEP here is obtained by calculating the CP radiation pattern using the term source method and weighting it by the laser's excitation coefficient (cf. §-III). Thus, the plot in figure 3(b) is the result of the product of the CP emission in air (without weighting) and the laser excitation that are both plotted in figure 4. As observed, the laser excitation field in figure 4 presents multiple peaks due to interferences which intensity quickly decreases and comes to a constant value. This behavior of the excitation field is due to the ZnO high absorption coefficient at λ=325nm which causes an exponential decrease of the amplitude of the transmitted field in the ZnO layer making the ZnO medium appear as an infinite medium. On the other hand, the plot of the CP emission in figure 4 also exhibits multiple interference peaks that will be further discussed in the following section. Finally, we remark, in figure 3, that the ZnO layer does not present any gain compared to the case of no ZnO layer ($d_{ZnO}=0$)

**C. CP radiation patterns**

We now analyze the radiation patterns of the polymer emission. Figure 5 represents a comparison between four of the polymer radiation patterns in the passive case for the TE mode. The top two graphs correspond to the CP radiation pattern in air while the bottom graphs represent the CP radiation pattern in Sapphire.

We now advance into explaining the origin of the emission peaks in air. The top-left radiation pattern (a) in Figure 5 corresponds to the case of $d_{ZnO}=134$nm, while the bottom-left radiation pattern (b) corresponds to $d_{ZnO}=200$nm. The two radiation patterns are quite similar; both patterns show a broad radiation pattern with a maximum at 0° and gradually decreasing with the increase of the emission angle ($\theta_{polymer}$). To understand the physics behind, one must realize that the radiation patterns are deeply affected by the density of optical modes [14]. A way to visualize the density of modes in a medium is by the reflectivity at the interfaces. More precisely, we can say that the radiation pattern is the result of the imbalance of the energy fluxes between the two interfaces.

In figure 6, we plot the reflectivity at both interfaces of the polymer layer for a ZnO thickness layer of 136 and 200nm. We remind that when the light is reflected from a higher index medium, which is the case with the CP/ZnO interface, the light encounters a change of phase of $\pi$, while when reflected from a lower index medium it encounters no change in phase. And so the reason of the minima and maxima in the luminescence becomes clear. In figure 6 we notice that while the reflectivity of the CP/Air interface is normally unchanged, the CP/ZnO interface reflectivity for $d_{ZnO}$=200nm is higher than that of $d_{ZnO}$=136nm. Since the reflected light wave on the CP/ZnO interface becomes anti-parallel due to the phase change, the augmentation of the reflectivity on the CP/ZnO interface then emphasizes the destruction of the emitted light in air and thus the intensity decreases. The shape of the radiation patterns is also explained by the reflectivity which decreases with the increase of the reflectivity on the CP/air interface while the reflectivity of the CP/ZnO interface is almost static all along the angles of the escape cone. Moreover, a detailed study of the reflectivity at the CP/ZnO interface for various ZnO thicknesses for the different values of $d_{ZnO}$, shows an inversely proportional relation between the CP/ZnO reflectivity and the CP emission as plotted in figure 4.

On the other hand, The radiation patterns in sapphire (Figure 5-a and 5-b) can be similarly explained . Radiation patterns are interesting in our study in order to determine the preferred collection angles and numerical apertures. For example, in air, it is better to collect at normal incidence with a wide numerical aperture.

### D. Parametrical study: Active ZnO layer

In this section, we study the effect of an active ZnO layer on the CP luminescence. Figure 7 shows a cross section (c, f) at $\theta_{excitation}$=0. Both cross sections have been normalized to $d_{ZnO}$=0. The left and right parts of figure 7 present respectively the collected TEP in air and in sapphire. In addition we show the TM (a, d) and TE (b, e) modes of the luminescence. We first remark that the CP emission is extremely TE polarized (TE:TM=10:1) and that the total emission spectra (c, f), which are the sum of the TE and TM modes, are loosely affected by the TM modes. Second, we notice that the emitted power in sapphire is stronger by a factor of 30 than the power emitted in air due to the narrow escape cone of the CP/air interface. Finally, concerning both collected power in air and in sapphire, we remark on a general enhancement of the CP luminescence (10~20x) with the increase of the ZnO layer thickness (Figure 7-c and 7-f). This enhancement is driven by the increase of the ZnO luminescence. Also, we notice repeating intensity oscillations that are the consequences of interference effects. These are followed by a saturation of the CP luminescence resulting from the saturation of the luminescence of the ZnO layer that is excited for a restricted layer thickness limited by the optical absorption of ZnO. The latter leads to an exponential decrease of the excitation field along the ZnO layer thickness. In Figure 7, the second emission peaks, which have a high intensity, exist at 136nm and 142nm ZnO layer thicknesses (indicated by the arrows) for the collected power in air and sapphire respectively.

### E.     ZnO emission spectrum

When excited under normal excitation densities, the ZnO usually emits light at a wavelength of 375nm due to free excitons. However, when exposed to high density excitation, many physical phenomena can be conjured (such as exciton-exciton scattering and electron-hole plasma) which results in a red-shift of the ZnO emission peak. In this part, we briefly discuss the influence of the wavelength red-shifting on the ZnO emission diagram. The later parameter was found to be an important and influencing parameter on the CP luminescence. Figure 8 presents the ZnO radiation patterns inside the ZnO layer itself. Only the radiation pattern of the light that is emitted towards the polymer is presented.  Graphs (a), (b), (c) and (d) correspond respectively to the emission wavelengths of 375, 380, 390 and 400nm. In order to fulfill the resonance (round-trip) condition, the ZnO layer thickness was varied for each of the graphs according to the relation $d_{ZnO} \approx \lambda/2n_{ZnO}$. This insures a constant slab reduced optical thickness. As shown in the next section, these thicknesses correspond to the maximum value of the ZnO excitation graph for each of the corresponding wavelengths (indicated by the arrows in Figure 9.b).

In the studied optical structure, the ZnO layer has the highest refractive index. And so, both escaping and guided modes can exist. Actually, even though guided modes are observed when light experiences total internal reflection; still the guided mode extends throughout the optical structure. So, while the escaping modes excite the CP layer through propagating light waves, guided modes can also excite the structure through evanescent light waves. Thus, both escaping and guided modes can contribute to the CP excitation. As a consequence, their contributions are considered in our calculations. Moreover, the total reflection condition at both of the ZnO interfaces is verified at incident angles superior than 45° for emission wavelengths of 375, 380 and 390nm; and at 53° for 400nm. At these angles, the light is confined inside the ZnO layer and is guided along the structure when the round-trip condition is verified.

Returning to Figure 8, first, we remark that a guided mode exists in all four cases. Second, we notice a remarkable enhancement of the magnitude of the guided mode from (a) to (d). In graph (d), the guided mode is very sharp and clear and dominates the radiation pattern. Although only the guided mode is visible in graph (d) because of scaling, the method takes into account all the kinds of modes including radiating and leaky modes as shown (indicated by the arrows) in the zoom of (d). Meanwhile, going from (c) to (a), the intensity of the guided mode drastically decreases till it becomes of the order of the escaping modes (a). This enhancement of the guided mode magnitude is due to the decreases of the ZnO absorption coefficient when the wavelength is red-shifted from 375 to 400nm. This red-shift exponentially increases the intensity of the light. Another effect of the lower losses is the decrease of the full width at half maximum (FWHM) of the guided mode as evidenced in figure 8 from (a) to (d). In fact, the FWHM and the losses are directly linked by the equation FWHM≈α with α being the attenuation coefficient.

In this section, we have discussed the effect of a red-shift of the ZnO emission wavelength. We introduce in the upcoming section the parametrical study on this aspect.

### F. Parametrical study: ZnO excitation

In this part, we further extend our study towards finding an optimal ZnO excitation. In figure 9 (a), we show the change of the CP TEP in air as a function of $d_{ZnO}$ as well as the value of the ZnO excitation for different ZnO emission wavelengths (figure 9.b). We first notice that both of the CP emission and the ZnO excitation gradually increase when passing from λ=375 to 400nm simply because of the decrease of the ZnO absorption which enhances the ZnO emission. In addition, we also notice the appearance of a dominant luminescence peaks for λ=390nm and 400nm. These peaks are due to the ZnO excitation. In fact, the ZnO excitation plotted in Figure 9 (b) along with the CP emission in Figure 9 (a) show that when increasing λ thus increasing the ZnO excitation, the behavior of the CP emission and the ZnO excitation becomes similar, which means that the ZnO excitation becomes predominant and controls the shape of the CP emission. The importance of the ZnO excitation and more specifically the excitation of a guided mode in the ZnO layer is further explained by the presence of the two maxima for $d_{ZnO} \approx \lambda/2n_{ZnO}$ which coincides with the cavity resonance condition.

## V. CONCLUSION

We have presented in this paper a theoretical study about the influence of a ZnO layer on the luminescence of a conjugated polymer ultra thin film. Both active and passive cases have been addressed. The study demonstrated that although the optical structure is simple yet it is promising for the enhancement of the luminescence and the sensing. We showed the interest of using an optically active ZnO for obtaining a drastic enhancement of the luminescence of the order of 20 times compared with the passive case. Furthermore, the study did not come on the subject of lasing which may occur at high pumping intensity accompanied by a shift of the emission wavelength as discussed in part E of section IV. The lasing along with the red-shift offers even higher prospects for luminescence enhancement. At higher wavelengths the ZnO layer exhibits lesser losses and with the gain obtained from the lasing process, one could expect a drastic increase of the ZnO layer excitation. This augmentation can be seen as further enhancement of the CP luminescence. We would also like to point out another important feature of the structure that is its simplicity; in fact based on planar thin films such a structure can be easily realized. Finally the concept of ultra thin film luminescence enhancement using an active substrate is not limited to the two materials reported in this paper.

**Acknowledgements**

The authors would like to thanks P. Le Barny, C. Galindo and L. Divay for providing the polymer thin film for ellipsometry. This work is supported by the ANR project, ULTRAFLU and the Champagne-Ardenne regional council.

**REFERENCES**

[1] H. Kautsky and H. DeBruijn, Naturwissenshaften **19**, 1043 (1931)

[2] Ma W., Hwang K.J., Lee V.H.L., Pharmaceutical Research **10**, 204-207 (1993)

[3] Brent S. Gaylord, Alan J. Heeger, and Guillermo C. Bazan, *J. Am. Chem. Soc.* **125**, 896–900 (2003)

[4] Bin Liu, Wang-Lin Yu, Jian Pei, Shao-Yong Liu, Yee-Hing Lai, and Wei Huang, Macromolecules, **34**, 7932–7940 (2001)

[5] Liming Dai, Prabhu Soundarrajan, and Taehyung Kim, Pure Appl. Chem. **74**, 1753–1772 (2002)

[6] D. Tyler McQuade, Anthony E. Pullen, and Timothy M. Swager, Chem. Rev., **100**, 2537–2574 (2000)

[7] Yong Cao, Ian D. Parker, Gang Yu, Chi Zhang & Alan J. Heeger, Nature **397**, 414-417 (1999)

[8] Lourdes Basabe-Desmonts, David N. Reinhoudt and Mercedes Crego-Calama, Chem. Soc. Rev. **36**, 993 – 1017 (2007)

[9] Jye-Shane Yang and Timothy M. Swager, J. Am. Chem. Soc. **120**, 11864–11873 (1998)

[10] Aimée Rose, Zhengguo Zhu, Conor F. Madigan, Timothy M. Swager & Vladimir Bulovic acute, Nature **434**, 876-879 (2005)

[11] Ü. Özgür, Ya. I. Alivov, C. Liu, A. Teke, M. A. Reshchikov, S. Doğan, V. Avrutin, S.-J. Cho, and H. Morkoç, J. Appl. Phys. **98**, 041301 **(**2005)

[12] W. Lukosz, Phys. Rev. B **22**, 3030–3038 (1980)

[13] Le Barny, Pierre L.; Obert, Edouard T.; Soyer, Francoise; Malval, Jean P.; Leray, Isabelle; Lemaitre, Noella; Pansu, Robert; Simic, Vesna; Doyle, Hugh; Redmond, Gareth; Loiseaux, Proceedings of SPIE-The International Society for Optical Engineering, 5990 (2005)

[14] R.P. Stanley, H. Benisty, M. Mayer, J. Opt. Soc. Am. A **15**, 1192–1201 (1998)

[15] K. G. Sullivan and D. G. Hall, J. Opt. Soc. Am. B **14**, 1149–1159 (1997)

[16] P. Yeh, Optical Waves in Layered Media (Wiley, New York, 1988)

[17] Yariv and Yeh, Photonics (Oxford University Press; Sixth Edition, 2006)

[18] K. Postava, H. Sueki, M. Aoyama, T. Yamaguchi, Ch. Ino, Y. Igasaki, and M. Horie, J. of Appl. Phys. **87**, 7820-7824 (2000)

[19] A. Yamilov, X. Wu, and H. Cao, J. of Appl. Phys. **98**, 103102 (2005)

**FIGURE CAPTIONS**

Figure 1. Studied optical structure (on the left) along with a representation of the source terms methods (on the right).

Figure 2. Parametrical study of the evolution of the conjugated polymer thin film (CP) total emitted power in air (above) and sapphire (below) as a function of the ZnO layer thickness and the angle of excitation. CP emission is excited both from the Laser and ZnO.

Figure 3. Total emitted power in air (a, b) and sapphire (c, d) for both TE (b, d) and TM (a, c) modes in case of a passive ZnO layer.

Figure 4. Plots of the conjugated polymer emission in air and the laser excitation field in the TE mode.

Figure 5. TE mode radiation patterns in air (a, b) and sapphire (c, d) for various ZnO layer thicknesses (70nm (a), 74nm (c), 100nm (b, d)).

Figure 6. Reflectivity coefficients on both of the conjugated polymer (CP) film interfaces (air, ZnO) as a function of the emission angle inside the CP ($\theta_{Polymer}$) for a ZnO layer thickness of 70nm (a) and 100nm (b).

Figure 7. Cross section at 0° of figures 2 and 3 (c, d) along with the TE (b, e) and TM (a, d) contributions on the luminescence. Graphs (c) and (f) are normalized to $d_{ZnO}=0$.

Figure 8. Radiation patterns of the ZnO inside the ZnO layer for 375 (a), 380 (b), 390 (c) and 400nm (d). The insets below graphs (c) and (d) respectively present a zoom on both graphs.

Figure 9. Evolution of the Total emitted power in air (a) and the ZnO excitation (b) as a function of the ZnO layer thickness and emission wavelength. Graphs are shifted by half an order of magnitude and normalized to $d_{ZnO}=0$.

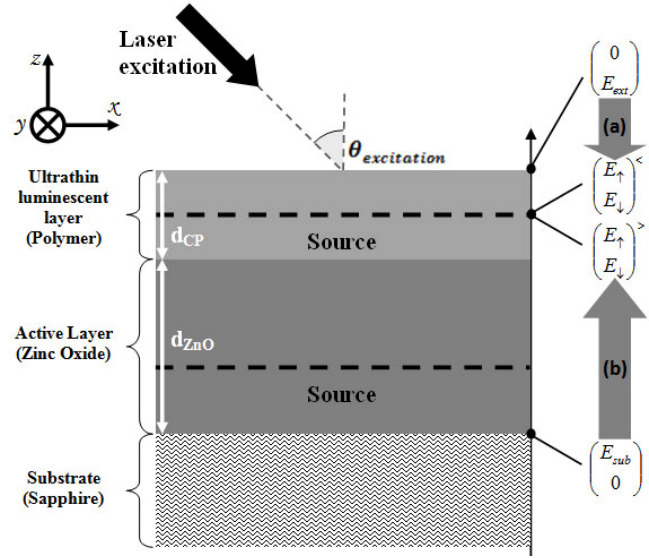

Figure 1

(a)

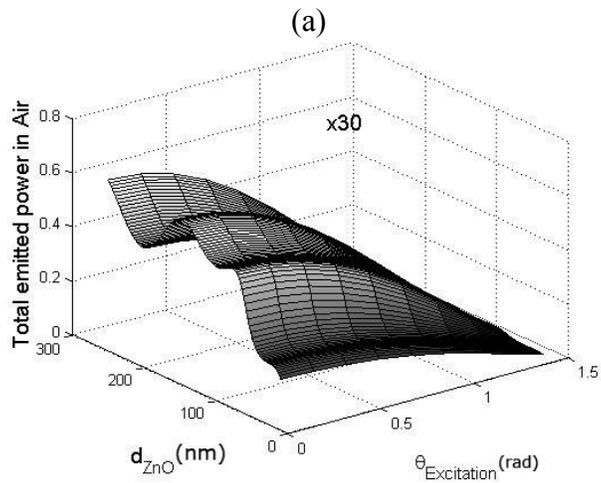

(b)

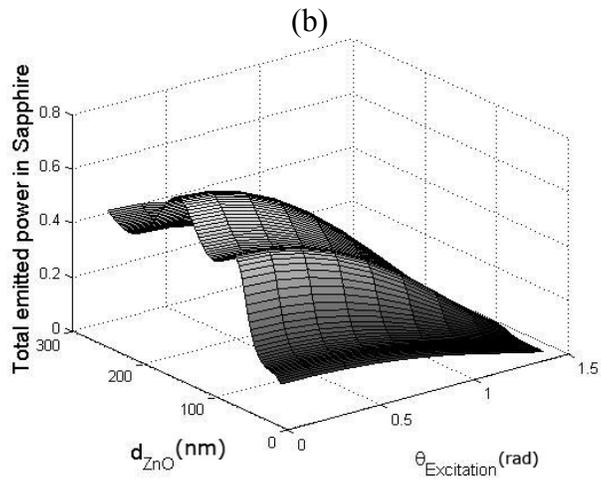

Figure 2

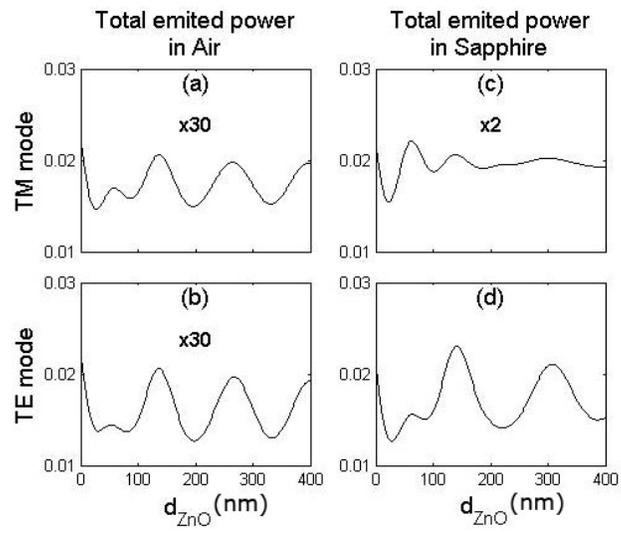

Figure 3

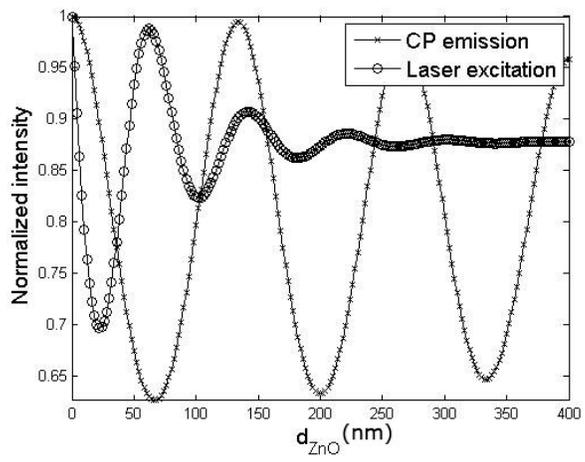

Figure 4

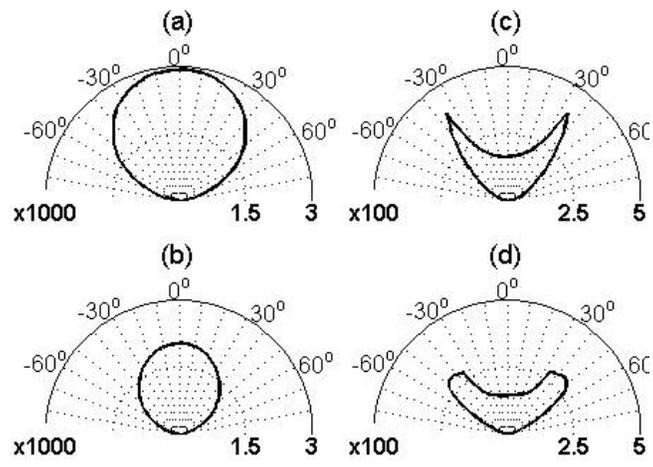

Figure 5

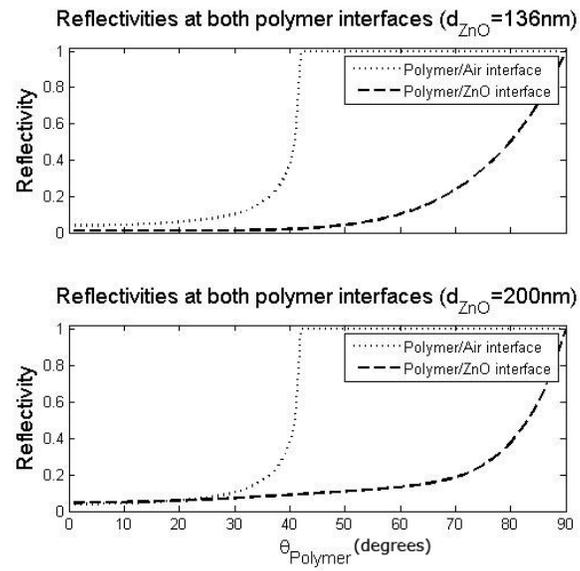

Figure 6

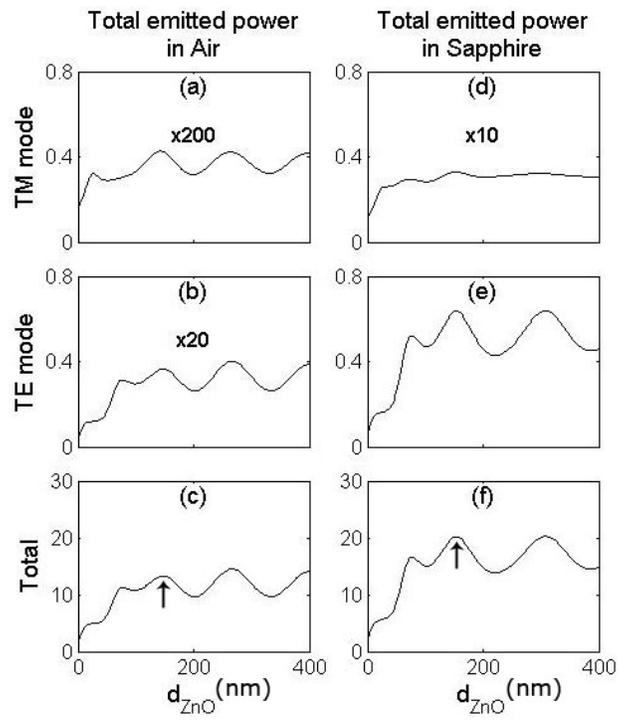

Figure 7

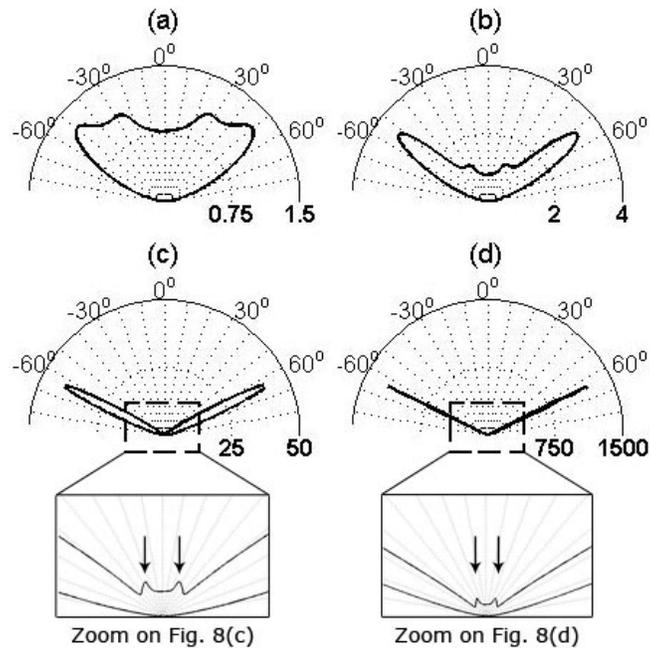

Figure 8

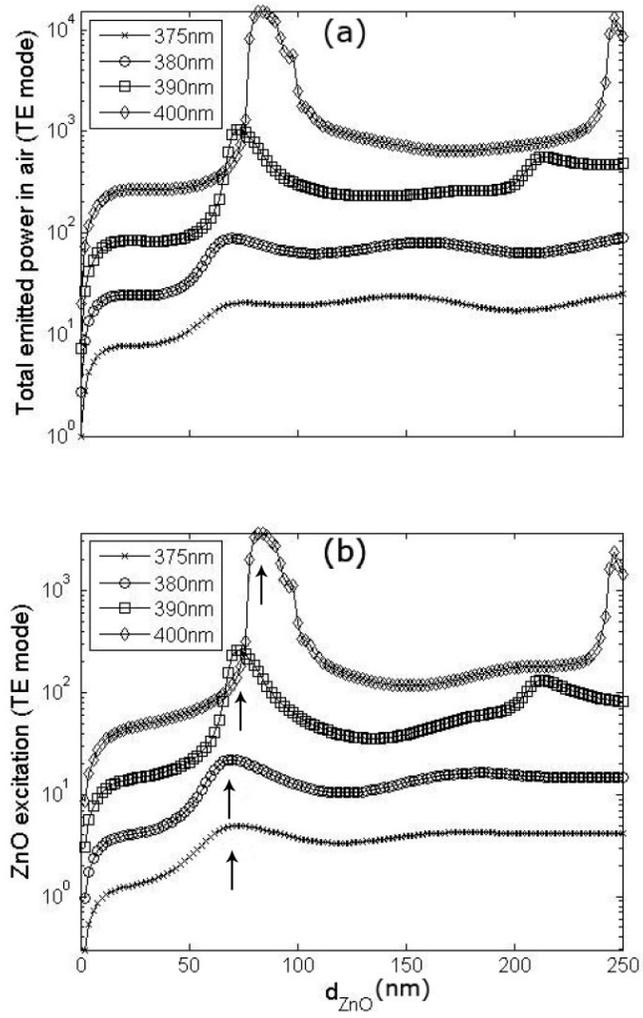

Figure 9

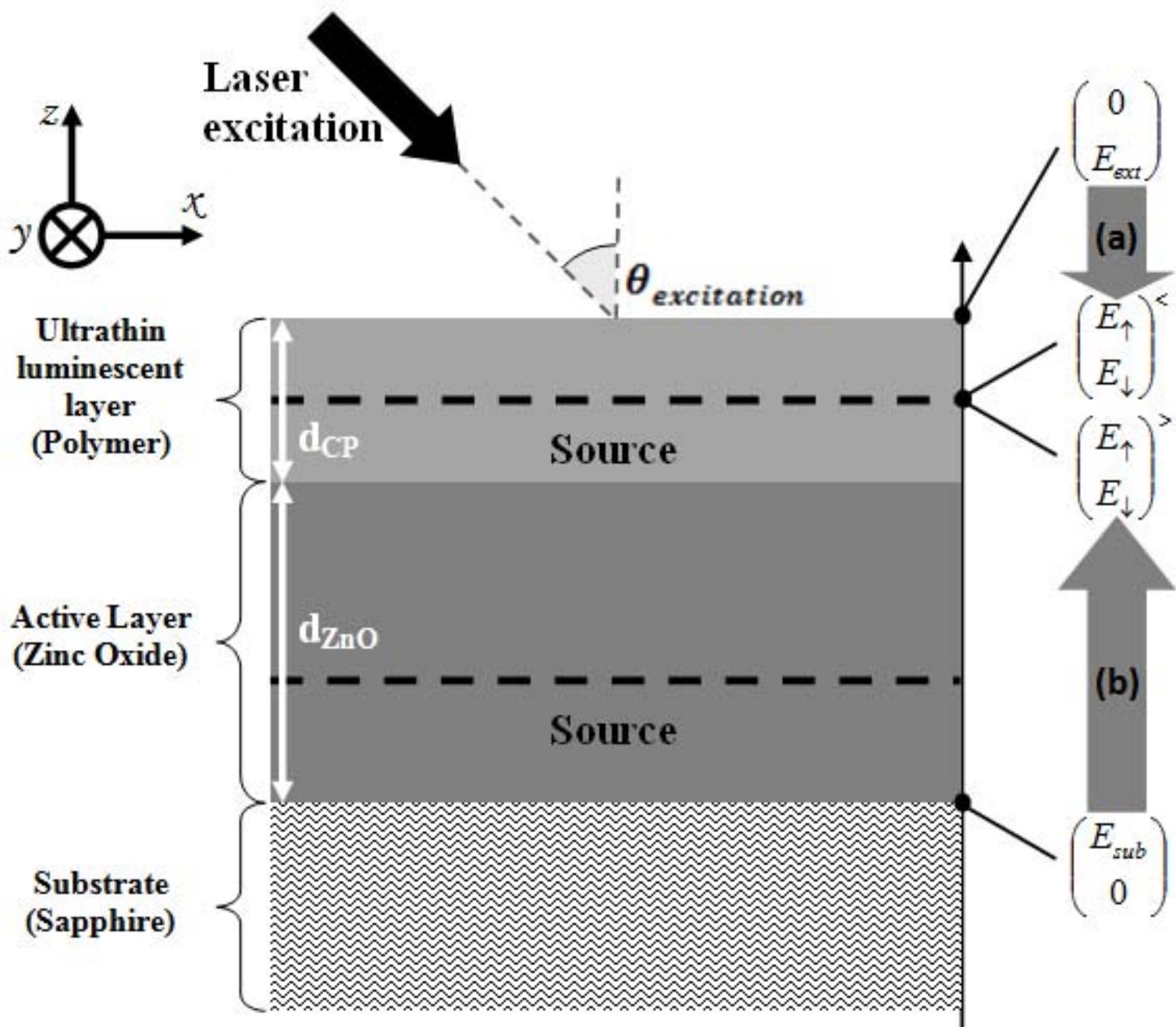

(a)

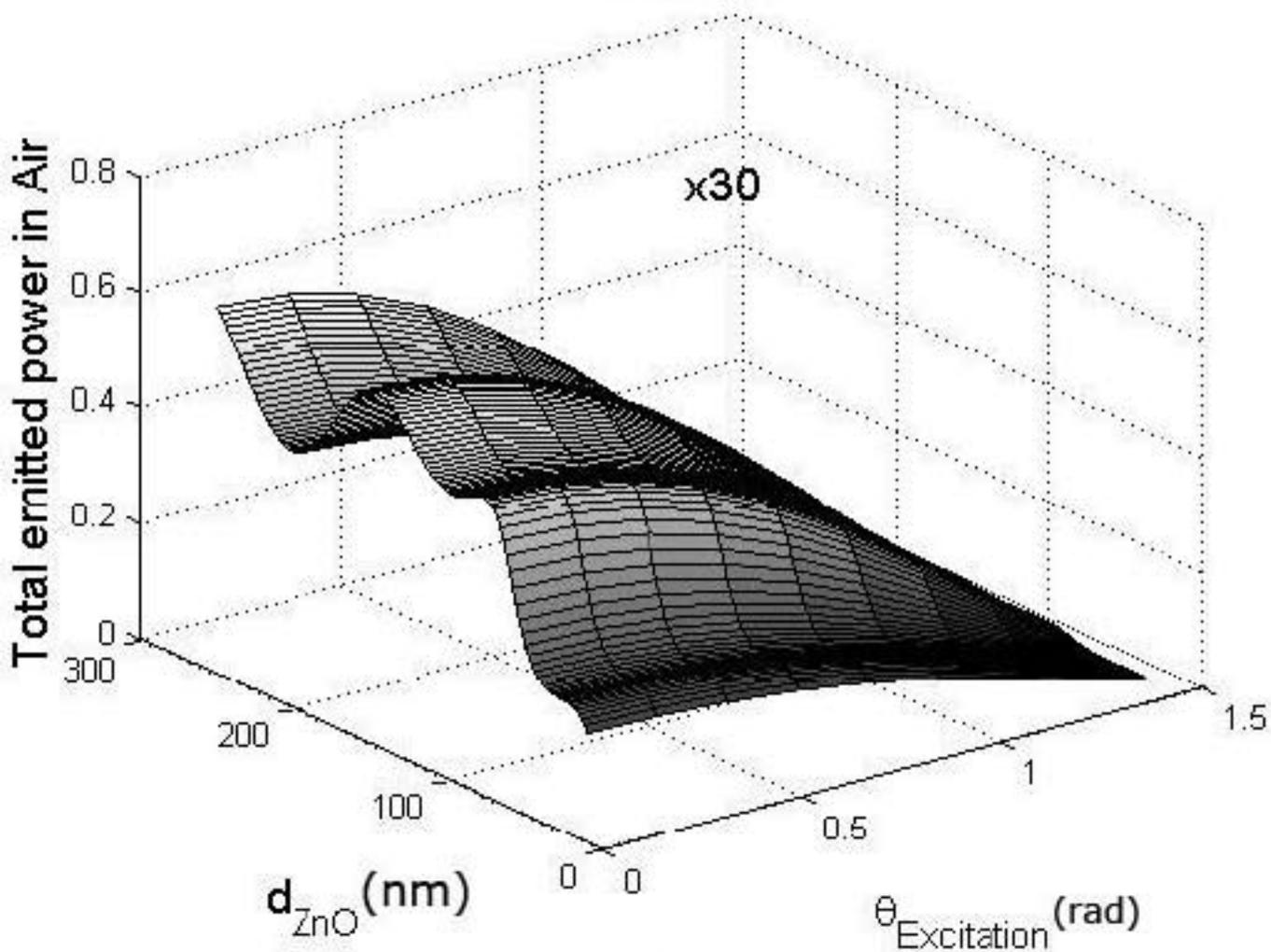

(b)

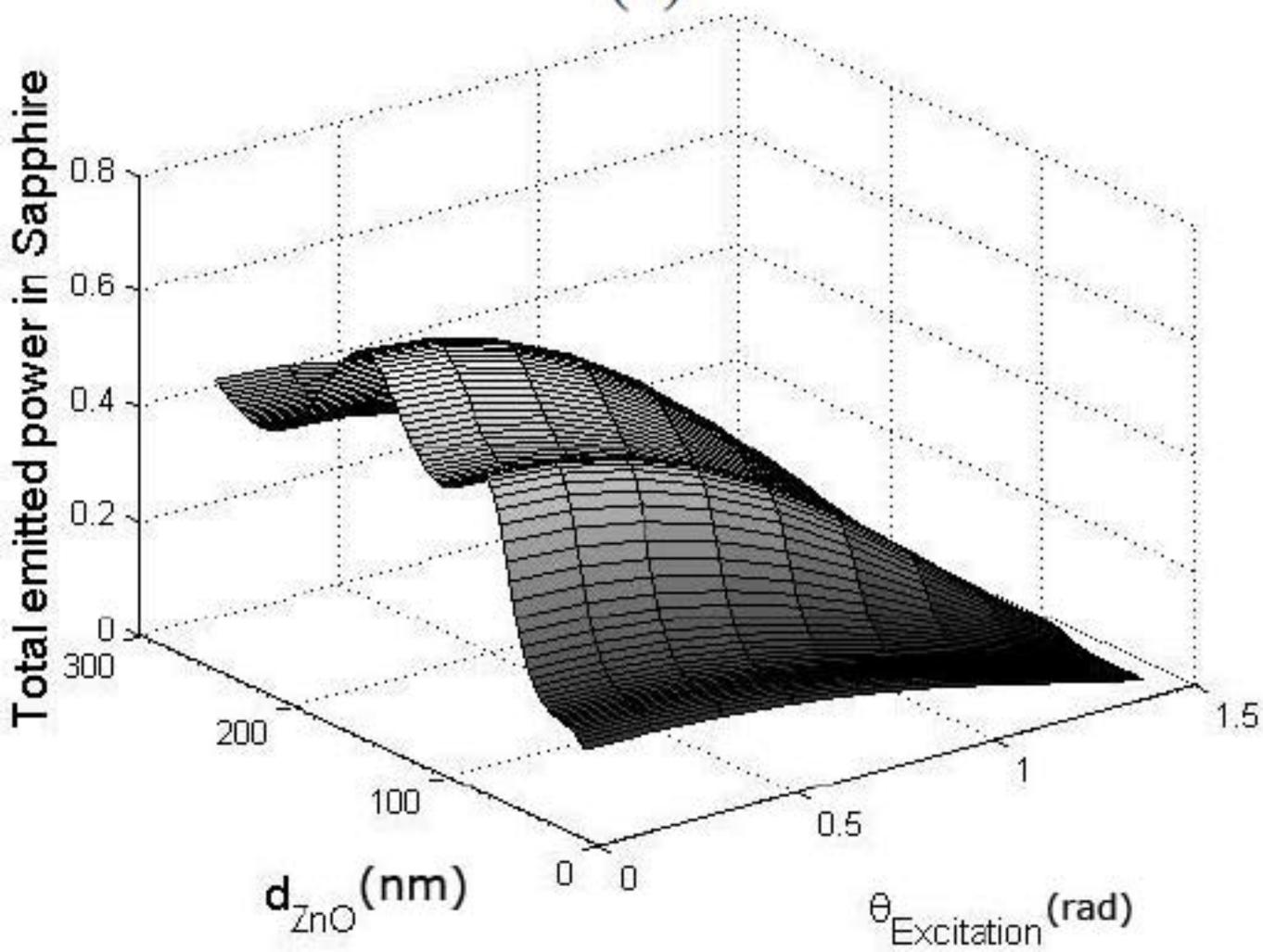

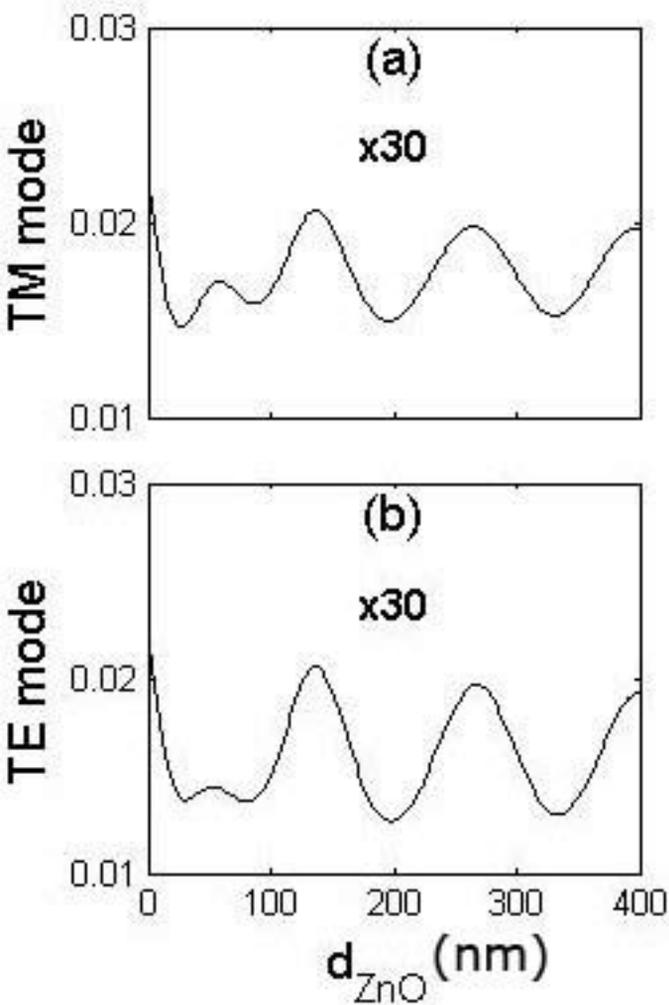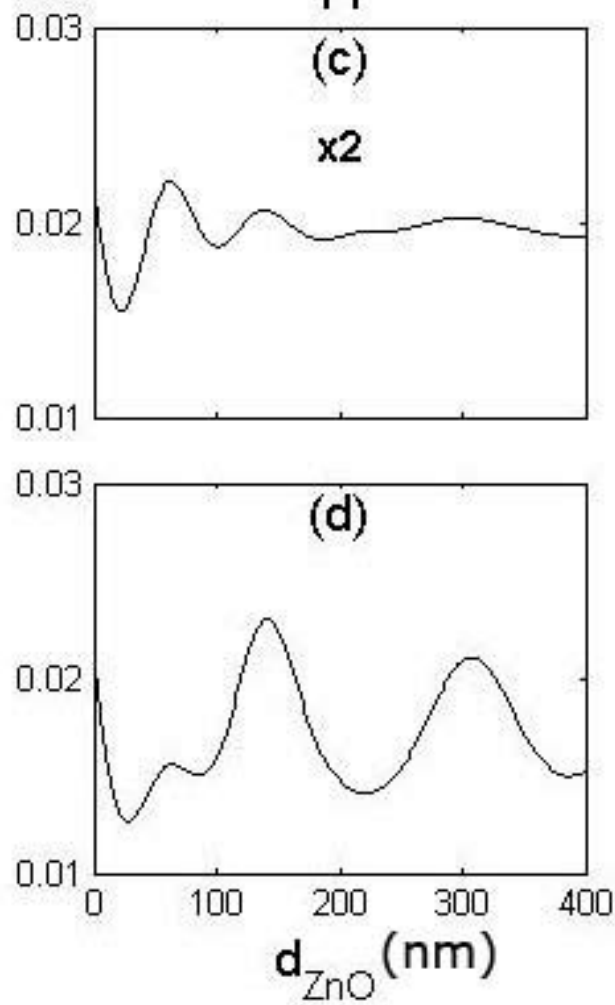

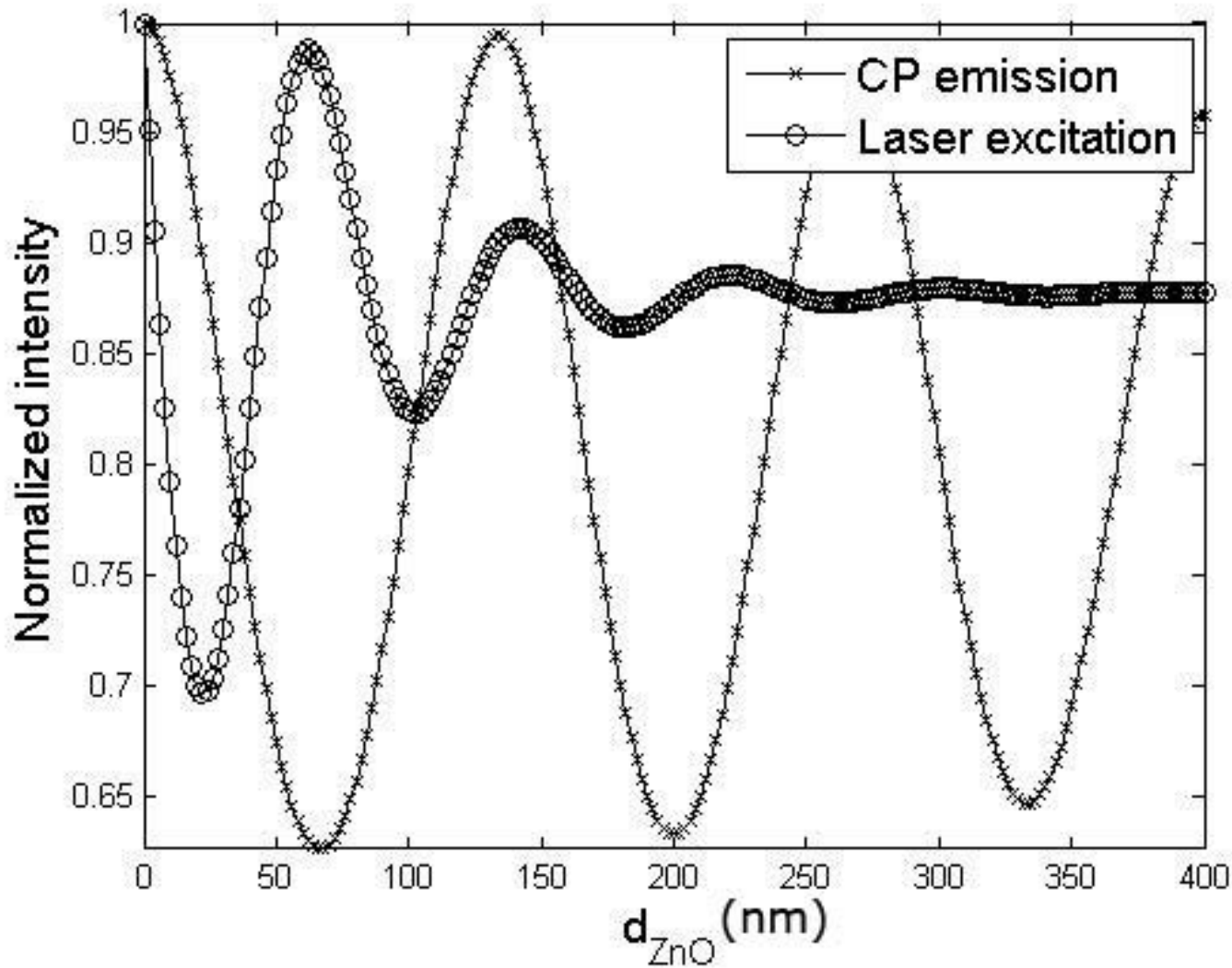

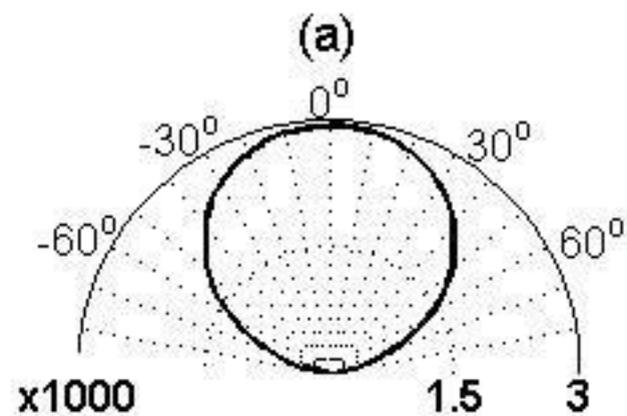 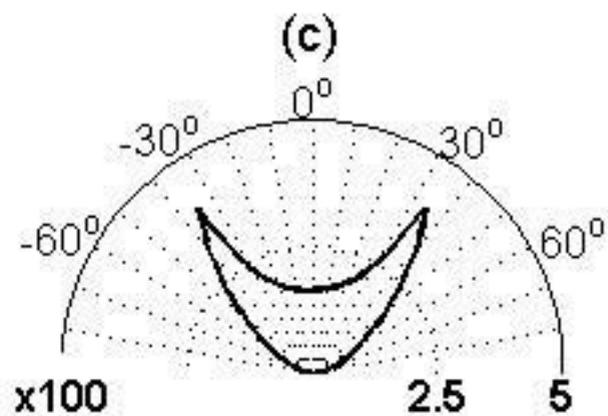
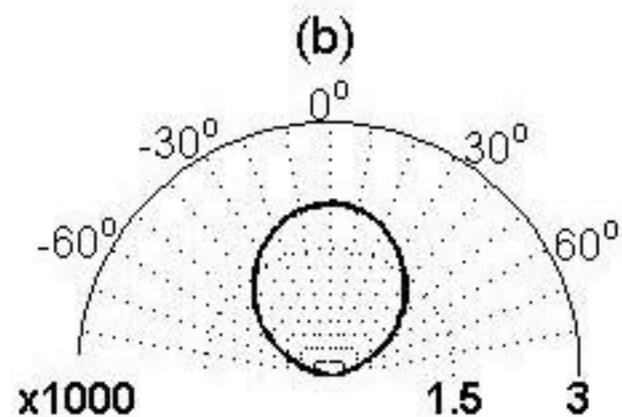 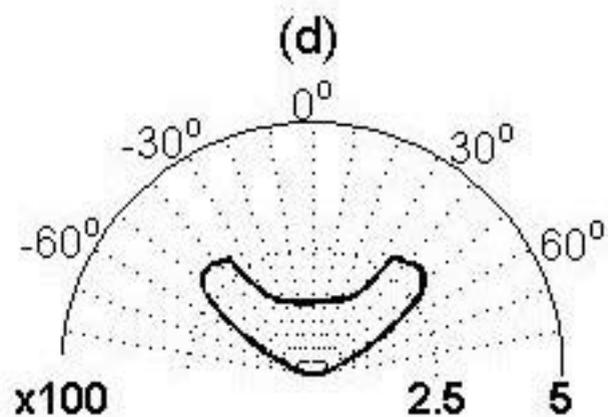

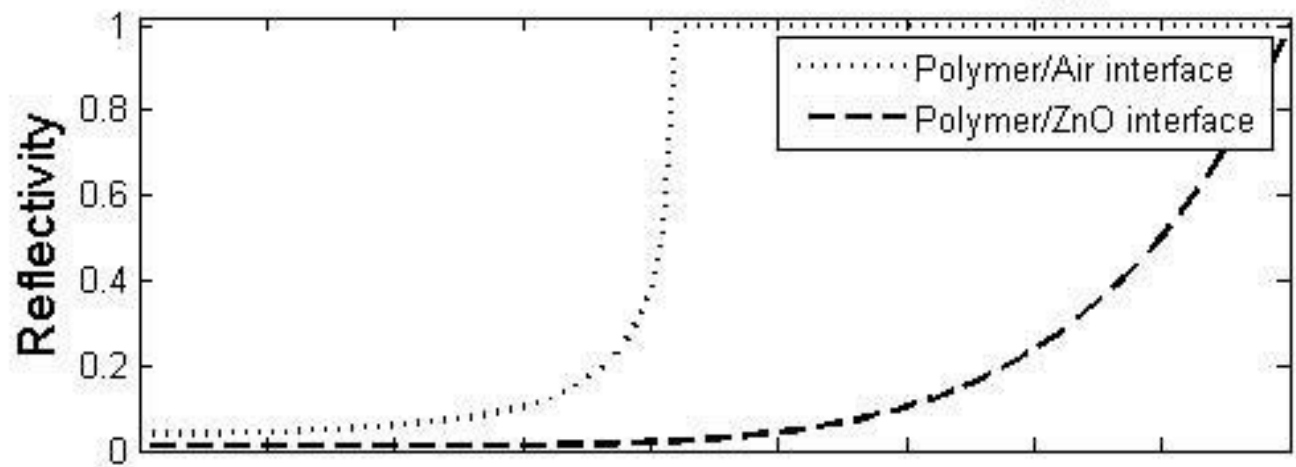
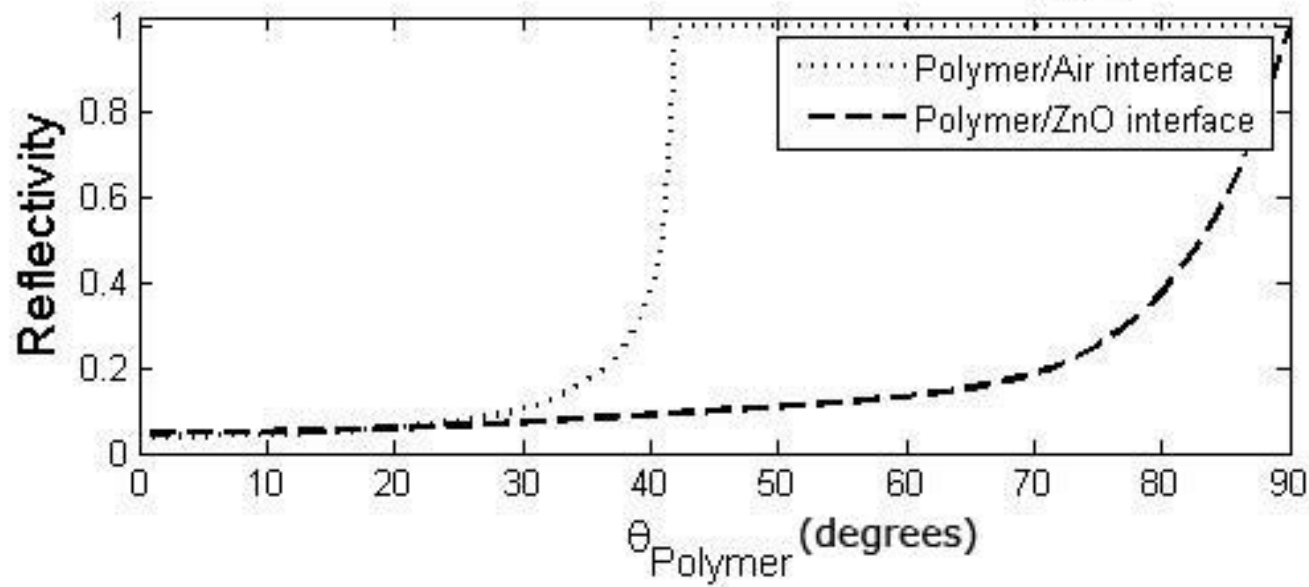

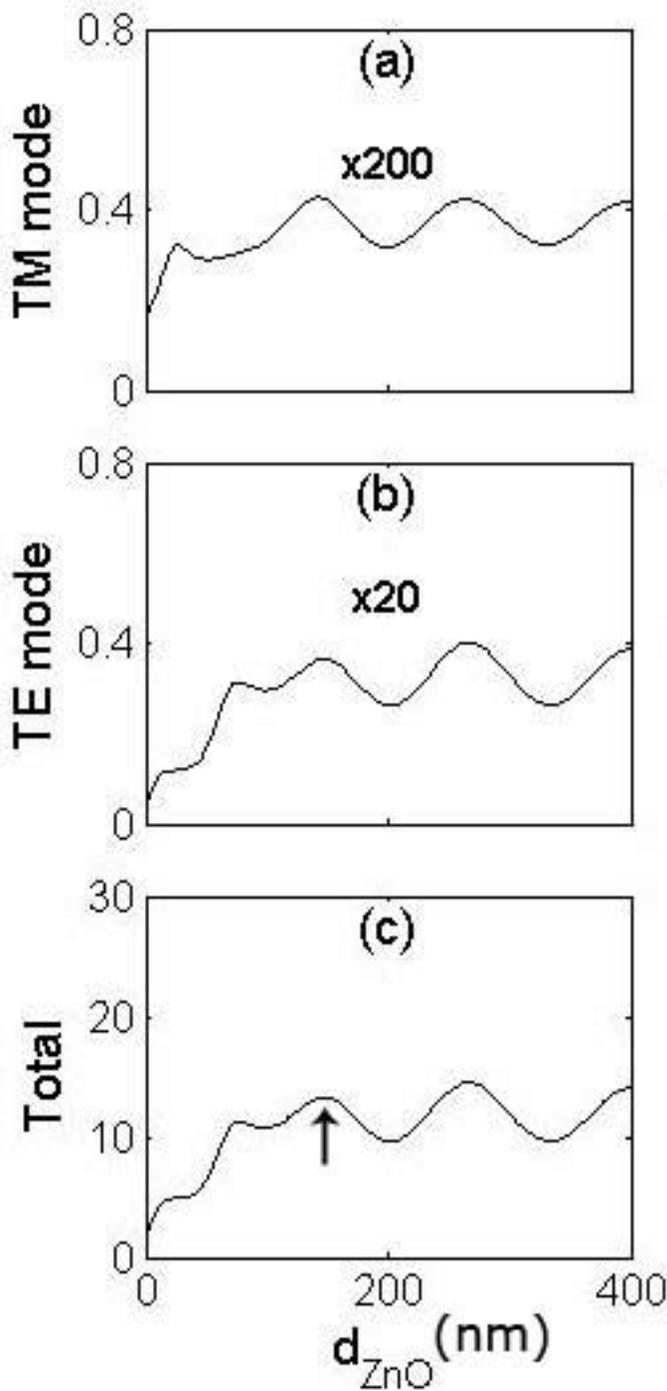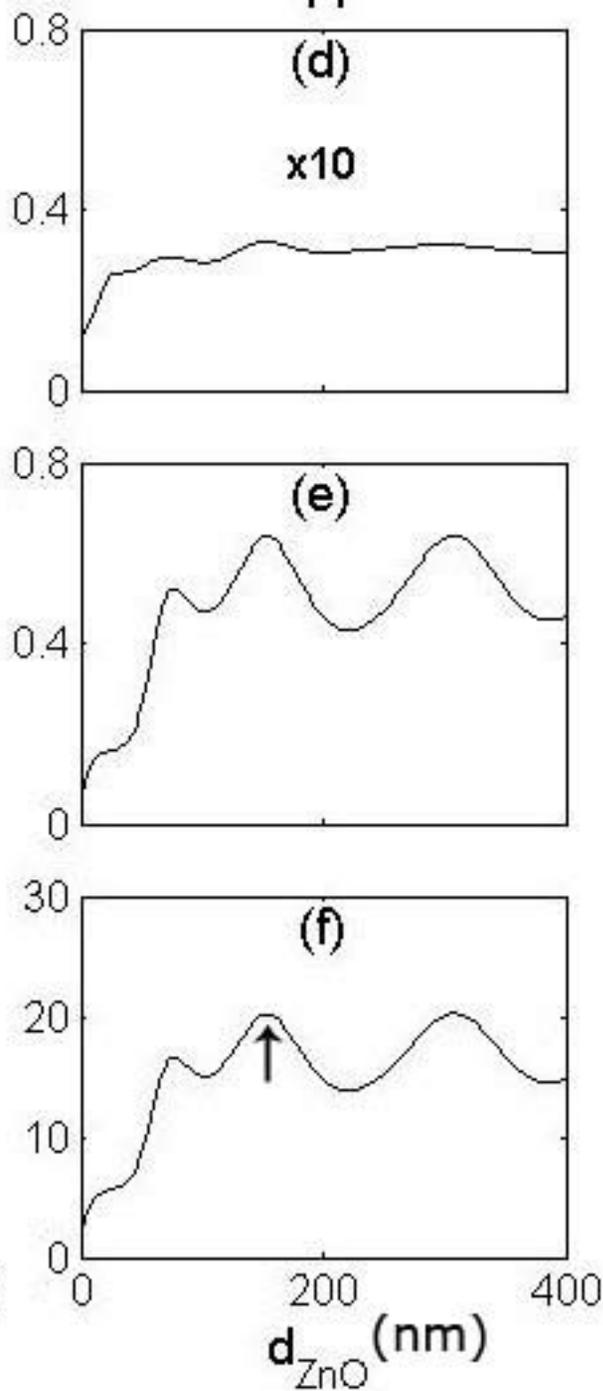

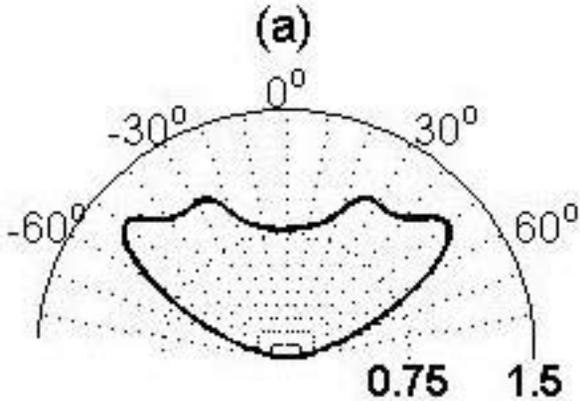
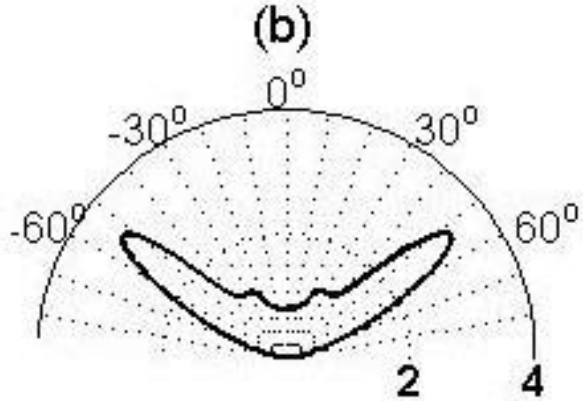
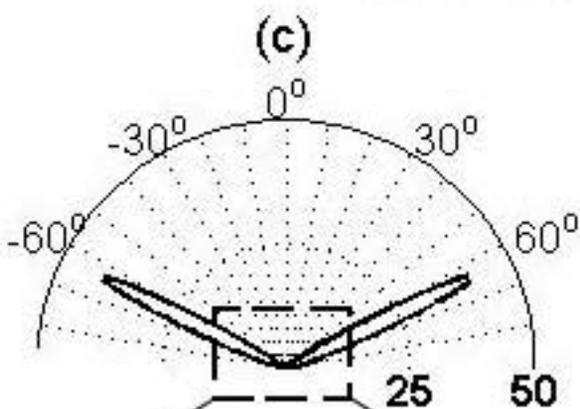
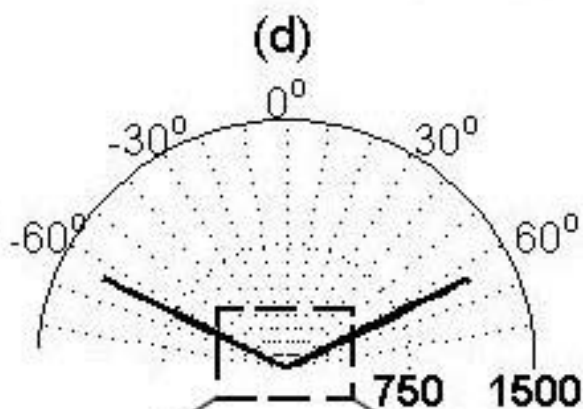
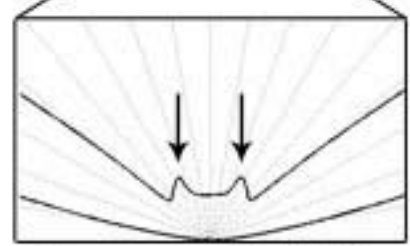
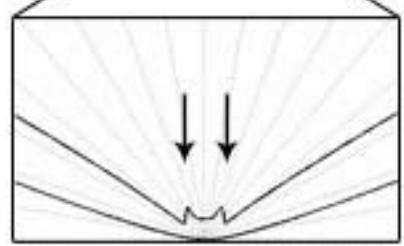

Zoom on Fig. 8(c)        Zoom on Fig. 8(d)

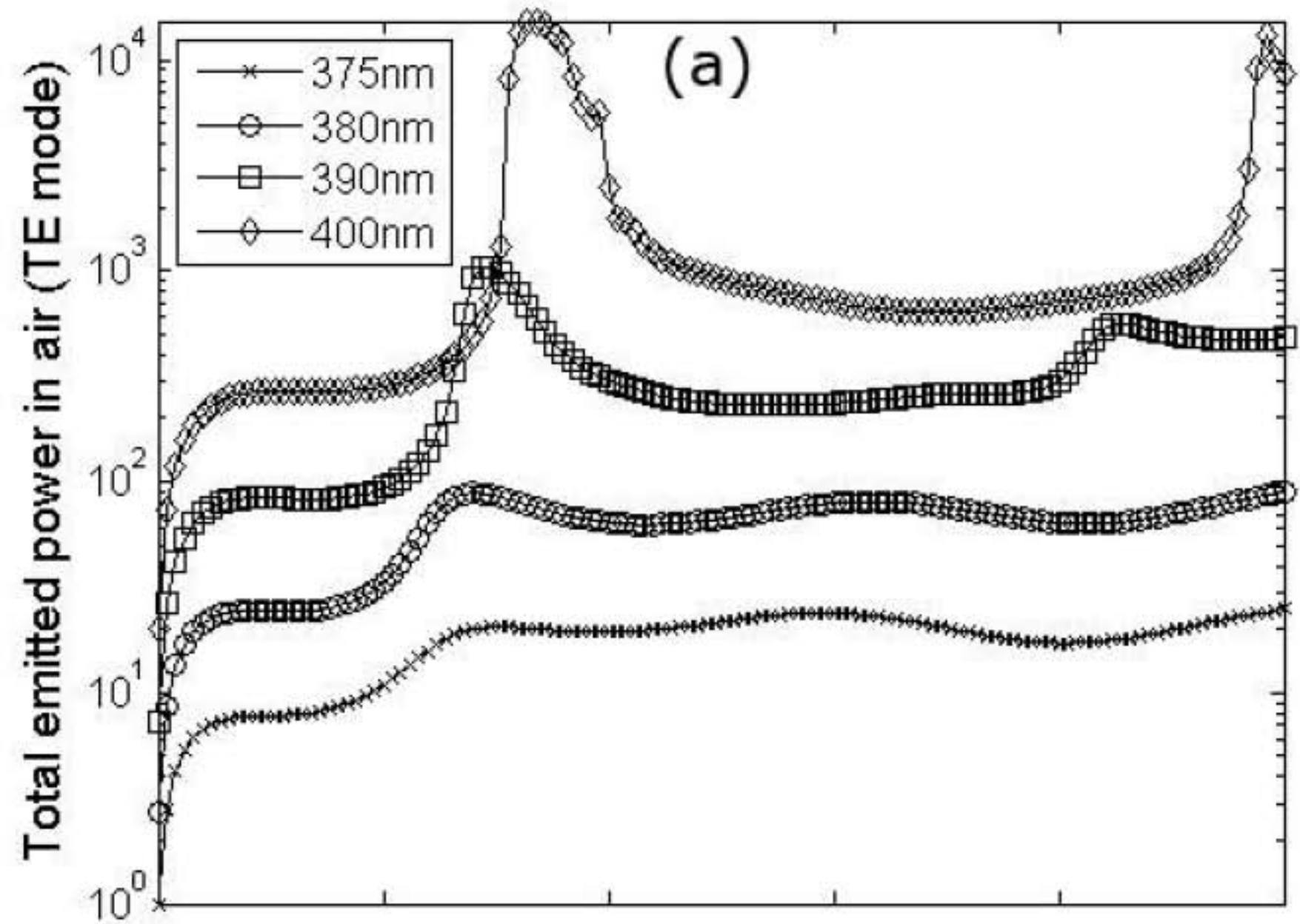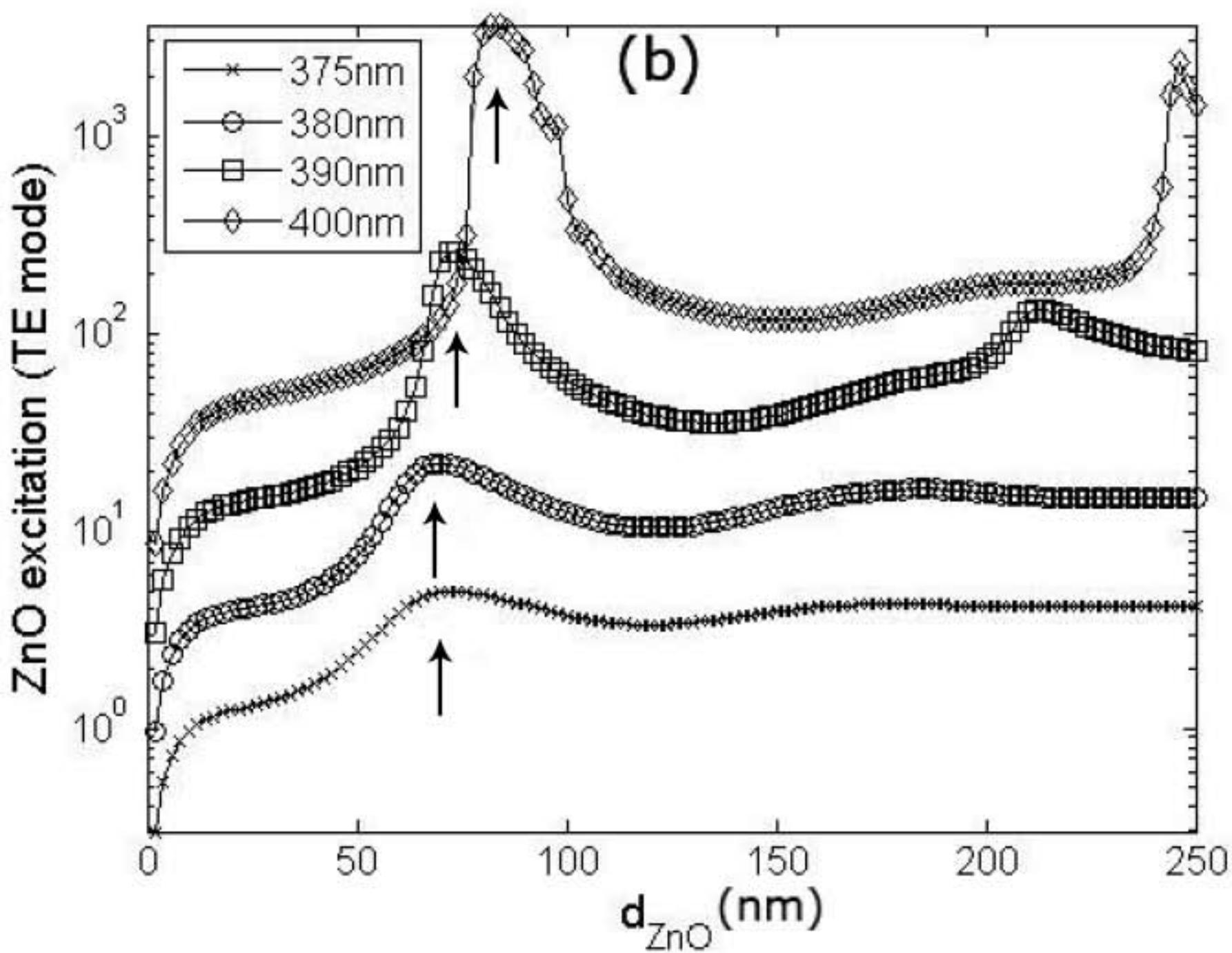